# High Precision Current Measurement for Power Converters

*M. Cerqueira Bastos*
CERN, Geneva, Switzerland

**Abstract**
The accurate measurement of power converter currents is essential to controlling and delivering stable and repeatable currents to magnets in particle accelerators. This paper reviews the most commonly used devices for the measurement of power converter currents and discusses test and calibration methods.

## 1    Introduction

Power converters for particle accelerators are required to deliver extremely accurate currents to the accelerator magnets. An essential element in guaranteeing the required performance is the current measurement device used for the control of the power converter current. For this reason, a lot of the developments within the field of current measurement for power converters over the last 50 years have been driven by the accelerator community and have happened in collaboration with the accelerator world. In the following sections the main current measuring devices available for power converter applications will be reviewed and discussed, including some of their advantages and disadvantages. Test and calibration methods will be discussed and their application to different accuracy requirements evaluated.

## 2    Metrology—Terms and definitions

There can be some confusion when using metrology terms in power converter applications for accelerators. Accelerator applications require repeatable and stable magnetic field conditions, and therefore power converters must provide repeatable and stable currents. The measurement devices used to measure these currents play a crucial part in obtaining this performance.

Current measuring devices are often specified by manufacturers and users in terms of 'accuracy' and 'precision'. It is therefore useful to clarify the use of these terms and some of the metrology vocabulary required to properly describe and characterize the performance of measurement devices.

According to the guide to the expression of uncertainty in measurement [1], accuracy is a qualitative concept referring to the closeness of agreement between a measurement and the true value of the measurand (or an accepted reference value). The quantitative expression of the accuracy (or lack of it) of a system is called uncertainty. Uncertainty is a non-negative parameter characterizing the quantity values attributed to a measurand. It is often expressed by a standard deviation. As for precision, it refers to the spread between measurements under the same conditions with no regard for the true value of the measurand [1].

The difference between precision and uncertainty is that uncertainty is given with regards to the true value of the measurand. The most common approaches to express uncertainty are:

- representing each component of uncertainty by a standard deviation;
- obtaining the combined uncertainty using the root-sum-of-squares method;
- multiplying the combined uncertainty by a coverage factor $k$, to increase the level of confidence.

A coverage factor of $k = 1$ means that 68.3% of the measurements are asserted to lie within the given uncertainty. A level of confidence of 95.5% corresponds to a coverage factor $k = 2$.

In accelerator applications, it is often useful to characterize a current measurement system's capabilities in terms of *gain* and *offset* errors, *linearity*, *repeatability*, *reproducibility* and *stability*.

Repeatability is extremely important in accelerator applications. Repeatability can be defined as the closeness of agreement between successive measurements carried out under the same conditions, while reproducibility refers to changing conditions [1]. The offset and gain errors refer to the systematic error at zero and full scale. Linearity describes a difference in the systematic errors throughout the measuring range, and stability can be defined as the change of measurement errors with time.

## 3    Current measuring devices

Table 1 presents a comparison of several well-known current measuring devices, including direct-current current-transducers (DCCTs) and current transformers (CTs) and some of their most important characteristics.

**Table 1:** Current measuring devices

|  | **DCCTs** | **Hall effect** | **CTs** | **Rogowsky** | **Shunts** |
|---|---|---|---|---|---|
| Principle | Zero flux detector | Hall effect | Faraday's law | Faraday's law | Ohm's law |
| Output | Voltage/current | Voltage/current | Voltage/current | Voltage | Voltage |
| Accuracy | Best devices can reach ppm uncertainty | Best devices can reach 0.1% uncertainty | Typically not better than 1% uncertainty | Typically percentage uncertainty, better is possible with digital integrators | PPM uncertainty for low currents, percentage for high currents |
| Ranges | Tens of A to tens of kA | Hundreds of mA to tens of kA | Tens of A to tens of kA | High currents up to 100 kA possible | From mA up to several kA |
| Bandwidth | kHz for higher DC currents, up to a couple of hundred kHz for lower DC currents | DC up to a couple of hundred kHz | Typically 50 Hz up to a few hundred kHz | A few Hz possible, up to MHz | Up to hundreds of kHz with coaxial assemblies |
| Isolation | Yes | Yes | Yes | Yes | No |
| Error sources | Magnetic (remanence, external fields, centring); Burden resistor (thermal settling, stability, linearity, temperature coefficient); Output amplifier (stability, noise, common mode rejection, temperature coefficient) | Magnetic; Burden resistor; Output amplifier; Hall sensor stability (temperature coefficient, piezoelectric effect) | Magnetic (remanence, external fields, centring, magnetizing current); Burden resistor | Magnetic integrator (offset stability, linearity, temperature coefficient) | Power coefficient; Temperature coefficient; Ageing; Thermal voltages |

As can be seen, DCCTs guarantee the best performance but they are also the most complex devices, with several possible error sources.

### 3.1 Current transformers

Current transformers (CT) are instrument transformers that produce, from an AC primary current, a proportional secondary current. The secondary winding is connected to a burden resistor to produce a measurable voltage signal.

The simplified equivalent circuit, referred to the secondary, of a current transformer [2] is shown in Fig. 1.

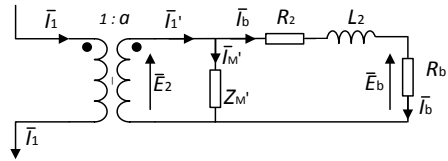

**Fig. 1:** Current transformer simplified equivalent circuit

The magnetizing current causes an amplitude and phase error in the CT. The secondary leakage impedance adds to the burden resistor, affecting the current distribution between $I_M$ and $I_b$. To improve accuracy, the magnetizing inductance must be maximized and leakage inductance must be minimized (high $\mu_r$, good winding distribution). It is important to remember that, as the CT approaches saturation, the magnetizing inductance decreases, increasing the error ratio.

Other sources of uncertainty include errors due to external magnetic fields. External fields move the CT closer to saturation and therefore a magnetic shielding should be used to minimize this effect.

Figure 2 shows the low frequency and high frequency models of the CT.

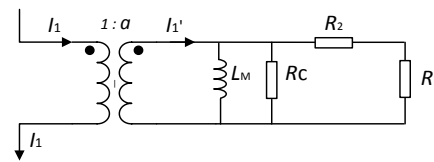

$$f_{LF} = \frac{\omega_{LF}}{2\pi} = \frac{R_C \| (R_b + R_2)}{2\pi L_M} \qquad (1)$$

(a)

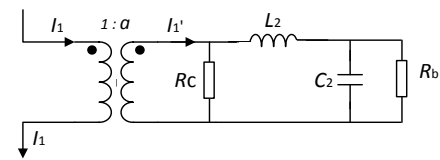

$$f_{p1} = \frac{\omega_{p1}}{2\pi} = \frac{1}{2\pi R_b C_2} \qquad f_{p2} = \frac{\omega_{p2}}{2\pi} = \frac{R_C}{2\pi L_2} \qquad (2)$$

(b)

**Fig. 2**: Current transformer (a) low and (b) high frequency simplified equivalent circuits

To extend the CT's low-frequency response the magnetizing inductance should be maximized. This means using high-permeability cores (e.g. silicon steel or nickel alloy). Limited low-frequency response causes droop in pulse measurement applications. The high cut-off frequency is mostly determined by leakage inductance and stray capacitance which, with some approximations, gives the origin of the two real poles shown in Fig. 2.

### 3.2 Hereward transformer

In the 1960s, H.G. Hereward proposed an active current transformer circuit that used electronic feedback to extend the low-frequency response of a CT and improve its accuracy (Fig. 3) [3]. In this device, an active circuit senses the voltage across the CT burden and uses a feedback loop to produce a compensation current that keeps the total flux in the cores at zero. The compensation current is a fractional image of the primary current. The effect of the feedback is equivalent to increasing the magnetizing inductance $L_M$ by $(A_L + 1)$, in which $A_L$ is the open loop gain of the sensing amplifier. This technique results in improved accuracy and extended low-frequency response.

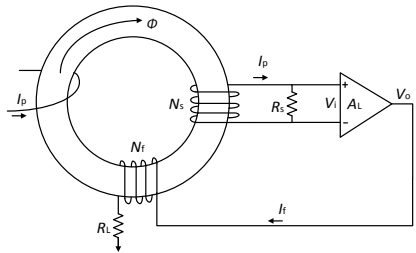

(a)

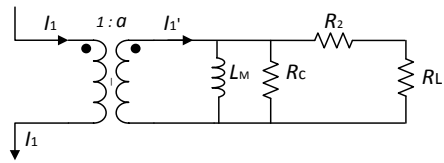

$$f_{LF} = \frac{\omega_{LF}}{2\pi} = \frac{R_C \| (R_L + R_2)}{2\pi L_M} \quad (3)$$

(b)

**Fig. 3:** Hereward transformer simplified equivalent circuit

### 3.3 DCCTs

In DCCTs this principle is taken further and combined with the magnetic modulator principle (used since the 1930s in fluxgate magnetometers) to provide an accurate measurement of currents ranging from DC to a few hundred kHz [4].

#### 3.3.1 Theory of operation

In a DCCT the primary current generates a magnetic flux that is seen by three cores (Fig. 4). A magnetic modulator drives two of the sensing cores in and out of saturation. The current peaks due to saturation of the cores are symmetrical if there is no DC flux in the cores and unequal if there is a DC flux in the cores. The current peak asymmetry due to any DC flux is measured and combined with the AC component measured by the third core. A control loop generates a compensation current that makes the total flux zero. This current is a fractional image of the primary current and can be measured with a burden resistor and an amplifier [5].

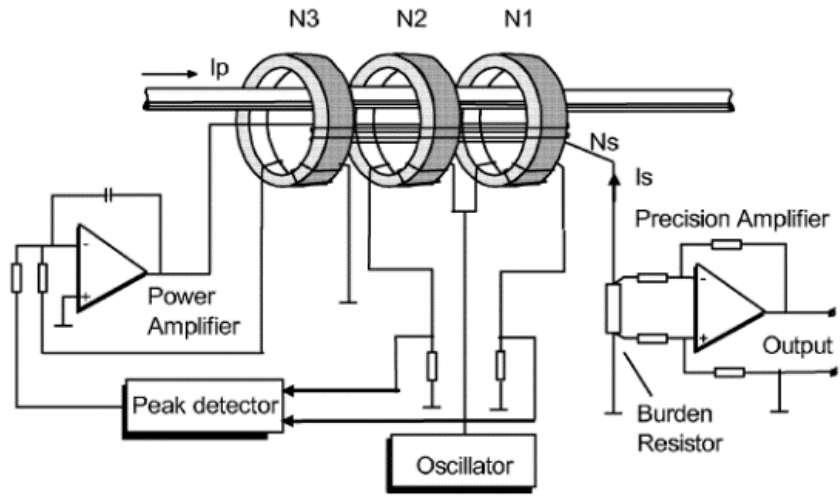

**Fig. 4:** DCCT theory of operation

### 3.3.2 *Sources of error in DCCTs*
Errors in DC measurements with DCCTs can come from different sources. The most common are related to the magnetic head, the burden resistor and the output amplifier:
- magnetic head: sensitivity to external magnetic fields, return bar, centring;
- burden resistor: gain error, settling at full scale (FS), stability at FS, linearity, gain temperature coefficient (TC);
- output difference amplifier circuit: offset error, settling at zero, stability at zero, offset and gain TC, noise, common mode rejection ratio (CMRR).

### 3.3.3 *Magnetic circuit*
The magnetic head of a DCCT plays an essential role in guaranteeing the ratio and offset accuracy in a DCCT. The magnetic circuit comprises high-permeability tape-wound magnetic cores with modulation windings. A magnetic shielding is used around the cores to protect them from external and leakage fields. The secondary windings are wound on top of the shielded assembly.

In addition to the factors mentioned in Section 3.3.2, magnetic remanence and modulation noise due to poor matching of sensing cores are other important factors to be considered in the design of the magnetic heads (Fig. 5).

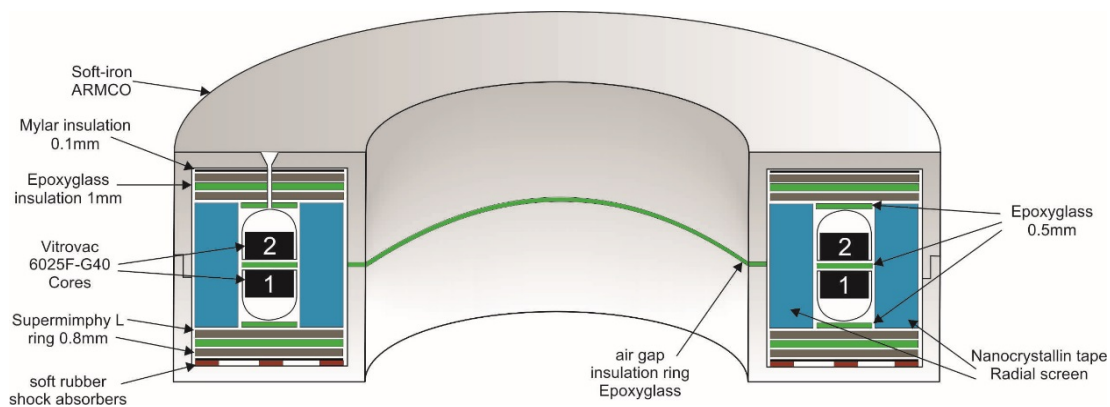

**Fig. 5:** DCCT magnetic head design

### 3.3.4 Burden resistor

The burden resistor is the most common source of uncertainty in DCCTs. Four-wire current sensing resistors are normally used. For power dissipation, a foil-based technology is a better choice than a wire. Common foil substrates are alumina and copper. A film deposited on a substrate is also a popular solution (thin film, thick film).

An important aspect to consider in the choice of resistors is that tolerance and stability do not always go together: processes that lead to tighter tolerances can result in degraded stability due to degraded power distribution and the creation of hotspots.

The main factors affecting resistor performance are related to their thermal behaviour. Well-known thermal effects upon resistance include the change of resistance with ambient temperature ($\Delta T$.TCR) and the change of resistance with internal temperature due to self-heating ($P.\theta R$.TCR). Less well-known effects include the power coefficient of resistance, which describes the transient effect of a change of resistance due to mechanical and thermal gradient phenomena linked to self-heating. Hysteresis under power cycling and humidity absorption/evaporation are other important factors that must be considered.

As mentioned above, resistor foil-based technologies are preferred, and bulk metal foil is now widely used in precision applications. It consists of a rolled metal foil (NiCr) bonded to a substrate, usually alumina. The foil/hard-epoxy/alumina combination is designed to give zero TCR to ambient changes of temperature: the foil TCR compensates for mechanical compression due to the substrate's lower thermal expansion coefficient. The resulting TCR is close to zero. This works well when temperature changes occur in all layers equally. However, with dissipation in the foil, thermal gradients are different, which results in an over-compression of the foil and the effective TCR turning more negative. The power coefficient of resistance describes this effect [6].

### 3.3.5 Output amplifier

The output amplifier of a DCCT is usually a difference amplifier circuit (Fig. 6). Some important points on the design of this stage are the gain resistor drift (matched networks are often used) and the common mode rejection, for which the matching of the gain ratios plays a crucial role. One should notice that the burden resistor affects the matching of the gain ratios, degrading the common mode rejection. This has to be taken in consideration in the choice of the ratio values. For the same reason, gain adjustment through potentiometers should be avoided and can be replaced by digital calibration.

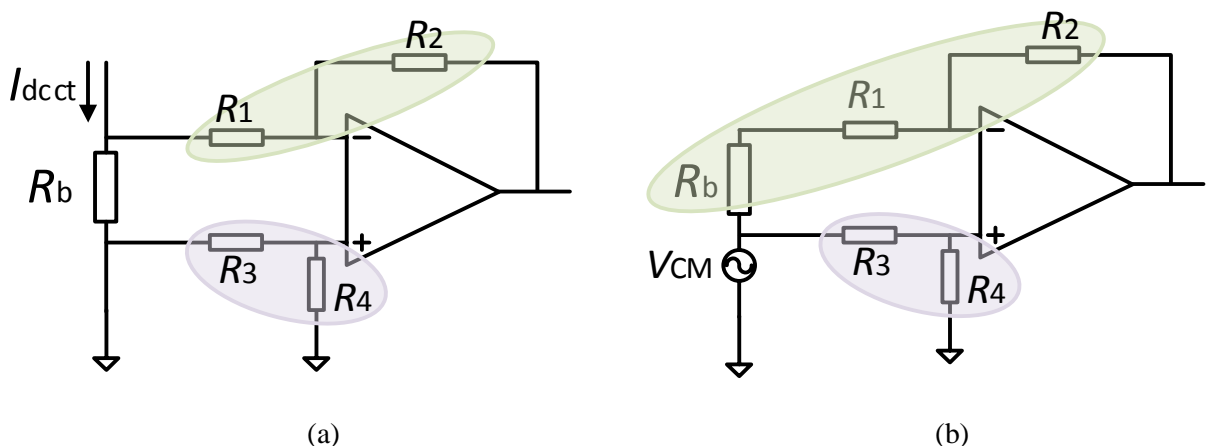

(a)                                                                 (b)

**Fig. 6:** DCCT output amplifier circuit. Matched ratio networks are typically used to implement the ratios R2/R1 and R4/R3 (a). The common mode rejection of the circuit is determined by the matching between the two ratios, which is affected by the burden resistor (b).

## 3.4 Hall effect transducers

Hall effect current transducers have been the transducers of choice for DC to AC current measurement applications, when percentage accuracy is required. They are cheap, simple and easy to use, with PCB and cable-mounted versions for a wide range of currents. There are basically two types of Hall effect probes:

– open loop: a Hall probe is placed in the air gap of a toroidal magnetic circuit. The magnetic flux generated by the primary current produces a Hall voltage in the probe, which is then amplified to produce the output signal;

– closed loop: Hall voltage is used in a closed loop to generate a compensating current, which is an image of the primary current.

When comparing these two types, closed loop models are preferred for higher accuracy applications, although limited to not better than 0.1% uncertainty. For open loop models, 1% uncertainty is typical. These limitations are mostly due to the stability of the Hall probe. Sensitivity to EMI can also be an issue. Concerning bandwidth: core geometry, thickness of laminations, core material and Hall chip all impact the bandwidth of open loop probes, typically not better than 50 kHz. Closed loop probes can go up to 200 kHz.

## 3.5 Selection of current measuring devices

The choice of a current measuring device for a given application depends on various factors: type of application (current range, bandwidth), required accuracy, required output signal (voltage, current), need for isolation from primary current circuit, reliability (mean time between failures (MTBF)), installation constraints, availability and cost.

For high current, high accuracy applications (e.g. >1 kA, <50 ppm), DCCTs are the devices of choice. Although voltage output devices are more common, current output devices are also available with secondary currents ranging typically from 1 A to 5 A. They are normally composed of a magnetic head and a separated electronic chassis. This chassis must be installed close to the converter control electronics in order to minimize output signal transmission distances. This type of DCCT usually guarantees very good reliability.

For medium current, high to medium accuracy (e.g. hundreds of amperes, <100 ppm), DCCTs are still the only devices able to deliver the required performance. In this case it might be easier and advantageous to use current output devices. Current output allows the designer to adapt the burden and amplifier choice and design to the required accuracy. In many cases, for these currents, electronics and head are integrated together and installed close to the power components of the power converter and far from the control electronics, this means a harsher environment and longer transmission distances for the output signal. In such cases, current output can be an advantage.

For low accuracy (e.g. percentage uncertainty) applications, current transformers and Hall effect transducers must be considered. Their simplicity and cost-effectiveness make them more attractive than the more complex and expensive DCCTs. Shunts are another current measurement solution for application where electrical isolation is not required.

A special mention for fast applications: if bandwidth is limited to a few hundred kHz, the solutions described above may still apply. In particular, DCCTs' small signal bandwidth can go up to a few hundred kHz. One thing to watch out for in DCCTs is the modulation voltage noise at the output of the DCCT and the voltage induced in the primary, which can be a problem in fast applications.

For bandwidths higher than some hundreds of kHz, Rogowsky coils and coaxial shunts should be considered, as long as accuracy requirements remain around the percent figure.

## 3.6 Test methods

There are several methods for evaluating current measuring devices and, in particular, DCCTs. The most important are discussed here.

In the reference device method, the primary current is measured both by the DUT and by a reference device, which are then compared (Fig. 7). According to ANSI/NCSL Z540.3-2006, the performance of the reference device must be at least four times better than the device under test (DUT).

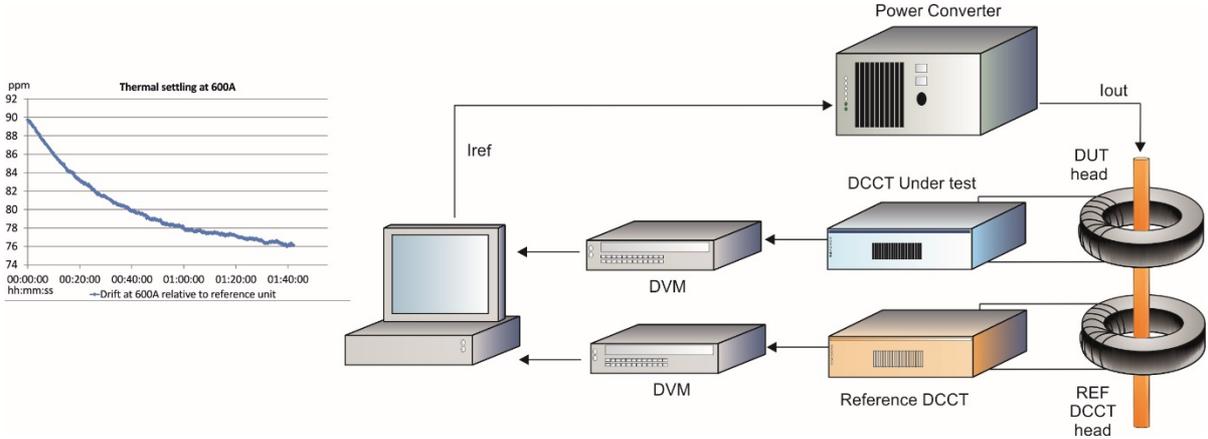

**Fig. 7:** The reference device test method

In the reference current method for testing DCCTs, a relatively small reference current is injected into an auxiliary winding with enough turns to simulate primary ampere turns (Fig. 8). The auxiliary winding can be permanent or temporary. The output of the DCCT is then measured with a precision digital voltmeter and the errors calculated.

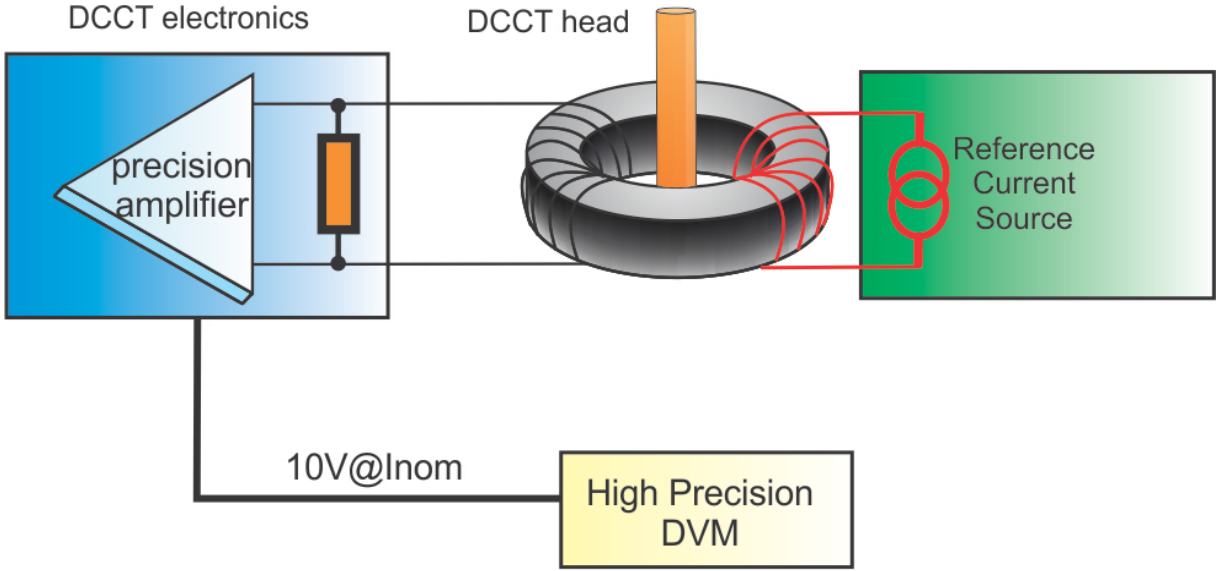

**Fig. 8:** The reference current test method

In DCCTs, a reference current can also be injected directly into the burden resistor in place of the compensation current (Fig. 9). The output of the DCCT is then measured with a precision digital voltmeter and the errors calculated. This test allows us to understand which errors are caused by the burden and output amplifier. This method must be used with care as common mode voltages between the burden resistor ground and the precision amplifier ground depend on the point of connection of the current source "low" terminal.

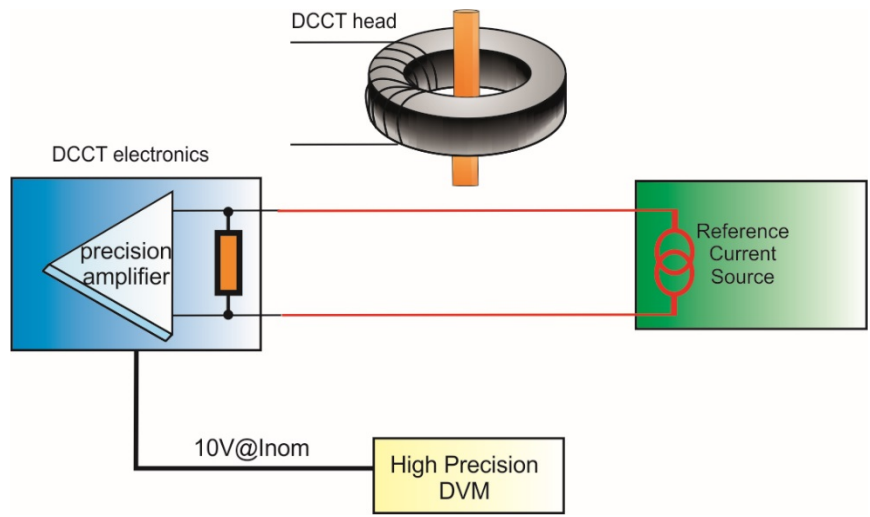

**Fig. 9:** The burden injection test method

On current output devices, the reference current can be compared 'back-to-back' with the DCCT current output as shown in Fig. 10. This test evaluates the quality of the current output of the DCCT, which is normally much better than the voltage output.

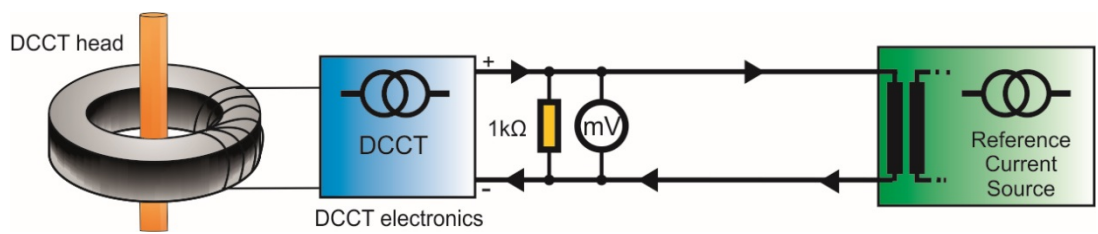

**Fig. 10:** The back-to-back test method

### 3.7 Test strategy

The test methods described above can be used for testing a current measuring device before installation and for calibration after installation. For high and very high accuracy applications, where <10 ppm measurement uncertainty is required, DCCTs are normally used. In these cases, intensive testing prior to installation is essential. These tests are performed in high-precision testbeds with specially chosen reference units or modified DCCTs. The test campaign can include installation and environmental tests such as centring, return bar influence, external magnetic field influence, temperature coefficient, EMC (voltage dips, burst test immunity, conducted noise), as well as performance tests such as gain, offset drift; settling at $I_{nom}$, linearity, noise, repeatability, reproducibility and settling at $I_{min}$.

In the most demanding applications these tests must be individually performed and the results properly tracked. For the less demanding applications, type-tests on a few units might be enough. In addition, integration tests should be performed upon installation of the DCCT in the power converters, to validate EMC and performance [7].

### 3.8 Calibration strategy

The first question to evaluate when deciding on a calibration strategy is knowing if a calibration strategy is really needed. Some of the issues to be considered when answering this question are:
– is long-term stability an important requirement in my application?
– what is the long-term drift of my current measuring devices?

- is there a need for tracking between different power converters?
- what is the impact of DCCT replacement? For example, replacing a DCCT that has drifted can cause a jump seen by the machine, and calibration can limit the size of this jump.

The calibration strategy must be adapted to the requirements of the application. For high- and very high-accuracy applications (e.g. <10 ppm relative accuracy, 50 ppm yearly drift) the calibration winding method and calibration against reference units are the best calibration methods. Both methods require specific calibration equipment, such as a suitable current source and suitable reference devices.

For medium accuracy applications (e.g. <100 ppm relative accuracy, 1000 ppm yearly drift) the comparison against a reference unit remains the best solution. The injection of a reference current into the burden resistor might be a more practical alternative for field calibrations where the installation of a reference unit is not possible. Again, these methods require either a suitable current source or suitable reference devices.

For low-accuracy applications with requirements in the order of percentage yearly drift, the need for calibration must be weighed against cost and effort, often resulting in a decision not to calibrate.